\documentclass[twocolumn,showpacs,preprintnumbers,amsmath,amssymb]{revtex4} 

\usepackage{epsfig} 
\usepackage{epsf} 
\usepackage{graphics} 
\usepackage{graphicx} 
\usepackage{amssymb}

\begin{document}

\setlength{\textheight}{23cm}
\setlength{\textwidth}{17cm}
\setlength{\oddsidemargin}{-.5cm}
\setlength{\topmargin}{-1cm}
\newcommand{\be}{\begin{equation}}
\newcommand{\ee}{\end{equation}}
\newcommand{\bq}{\begin{eqnarray}}
\newcommand{\eq}{\end{eqnarray}}

\title {A Lorentz-violating SO(3) model: discussing the unitarity, causality and 
the 't Hooft-Polyakov monopoles} 
\date{\today} 
\author{A. P. Ba\^eta Scarpelli$^{a,c}$ } \email []{scarp@fisica.ufmg.br} 
\author{J. A. Helay\"el-Neto$^{b,c}$} \email[]{helayel@cbpf.br}  

\affiliation{$^{a}$Universidade Federal de Minas Gerais \\
Departamento de F\'{\i}sica - ICEx \\
P.O. BOX 702, 30.161-970, Belo Horizonte MG - Brazil}

\affiliation{$^{b}$ CBPF, Centro Brasileiro de Pesquisas F\'{\i}sicas \\
Rua Xavier Sigaud 150, 22290-180 Urca \\
Rio de Janeiro, RJ, Brazil  } 

\affiliation{$^{c}$Grupo de F\'{\i}sica Te\'orica Jos\'e Leite Lopes \\
Petr\'opolis, RJ, Brasil.}

\begin{abstract}
\noindent
In this paper, we extend the analysis of the Lorentz-violating Quantum Electrodynamics 
to the non-Abelian case: an SO(3) Yang-Mills Lagrangian with the addition of the non-Abelian 
Chern-Simons-type term. We consider the spontaneous symmetry breaking of the model and 
inspect its spectrum in order to check if unitarity and causality are respected. 
An analysis of the topological structure is also carried out and we show that a 't Hooft-Polyakov 
solution for monopoles is still present.
\end{abstract}

\maketitle 

\section{Introduction}

The possibility of violation of Lorentz and CPT symmetries has been vastly investigated 
over the latest years \cite{1} - \cite{31}. Although they are in the basis of the modern Quantum 
Field Theory construction and despite the fact these symmetries are respected by the Standard Model 
for the elementary particles, one speculates if this scenario is only an effective theoretical 
description of a low-energy regime. In this case, these symmetries could be violated at 
an energy near the Planck scale.

The recent papers treat mainly the inclusion, in the gauge sector of the traditional 
Quantum Electrodynamics (QED), of a Chern-Simons like term
\be
\Sigma _{CS}=-\frac{1}{4}\int d^4x \epsilon ^{\mu \nu \alpha \beta }c_{\mu}
F_{\nu \alpha }A_{\beta},  
\ee
in which $c_\mu$ is a constant four-vector which selects a special space-time direction. Such a term 
would originate an optical activity for the vacuum. This optical activity is presented like 
a possible explanation for the pattern observed in the detection of ultra-high energy 
cosmic rays ($E>E_{GZK} \sim 4 \cdot 10^{19}KeV$) \cite{25}. Besides that, measurements of radio 
emission of distant galaxies and quasars detected that the polarization vectors of these 
radiations are not randomly oriented, as we would expect.  
Another interesting discussion has been raised from the investigation of the possibility that 
this term be radiatively generated from the fermionic sector of ordinary QED, when an 
axial term, $b_\mu \bar \psi \gamma^\mu \gamma^5 \psi$, which also violates Lorentz and 
CPT, is included \cite{11}-\cite{ebert}.  

In the paper \cite{24}, the quantization consistency of an Abelian theory which 
incorporates the term (1), with spontaneous symmetry breaking (SSB), 
has been contemplated. It 
was also studied its topological structure, with the 
discussion on vortex configurations affected by the 
direction of $c^\mu$ in the space-time. Interesting peculiarities were observed in the way 
the degrees of freedom are distributed amongst the physical modes of the theory after 
spontaneous symmetry breaking. Remarkable properties are also 
pointed out in the study of vortex configurations. In this work,  we consider 
a non-Abelian version of this Lagrangian with violation of Lorentz and CPT symmetries. 
Indeed, we treat a Yang-Mills Lagrangian with internal symmetry SO(3), with the inclusion of a 
non-Abelian Chern-Simons-type term:
\be
\label{2}
\Sigma_{CS}^{NA}=-\frac 14 \int d^4x \epsilon ^{\mu \nu \alpha \beta }c_\mu 
\left( G_{\nu \alpha}^a A_\beta^a - \frac e3 
f ^{abc}A_\nu^a  A_\alpha^b A_\beta^c \right), 
\ee  
where, like in the Abelian case, $c^\mu$ is a constant four-vector and the latin indices 
refer to the isospin space.
In this context, the SSB is discussed and the quantization consistency is analyzed. This study 
is carried out by the investigation if unitarity and causality are maintained in the gauge field propagators. The 
discussion is performed at the tree-level, and we focus on the analysis 
of the residue matrix associated with each pole of the propagator. 

We also analyze the topological structure of the theory. As we know, in a conventional
situation, when the 
symmetry group is non-Abelian and spontaneous symmetry breaking takes place, the 
field equations yield a solution which corresponds to a magnetic charge (the 
't Hooft-Polyakov monopole). Here, we investigate if this solution is still present 
and how the Lorentz-breaking term manifests itself in this context.

The paper is organized as follows: in the Second Section, we discuss the SSB and the 
quantization consistency is investigated from the poles of the gauge field propagator; 
Section 3 is devoted to the analysis of the field equations and their topological 
structure; and, finally, in the Section 4, we discuss the results and present 
our General Conclusions.

\section{The non-Abelian Lorentz-breaking gauge-Higgs model}

We begin our analysis by considering the action
\be
\Sigma=\int d^4x \left \{-\frac 14 G_{\mu \nu}^a G^{\mu \nu a} 
+ \frac 12 (D_\mu \varphi)^2 - V(\varphi) \right \}
+ \Sigma_{CS}^{NA},
\ee
where
\be
G_{\mu \nu}^a=\partial_\mu A_\nu^a - \partial_\nu A_\mu^a 
+e f^{abc}A_\mu^b A_\nu^c,  
\ee
\be
(D_\mu \varphi)^a= \partial_\mu \varphi^a + e f^{abc} A_\mu^b \varphi^c
\ee
and
\be
V(\varphi)=\frac 14 \lambda (\varphi^2-a^2)^2.
\ee
As already mentioned, $A_\mu^a$ and $\varphi^a$ are components of vectors in 
the internal space referring to 
the SO(3) symmetry. The $\vec \varphi$-field has a non-vanishing vacuum 
expectation value, $| \langle 0|\vec \varphi|0 \rangle |=a \hat u$, and the SO(3) gauge 
symmetry is spontaneously broken. We choose the vacuum that points in the internal 
$z$-direction,
\be
\vec \varphi_0=a\hat z.
\ee
Now we can make use of the fact that we have a local symmetry and choose the so-called 
unitary gauge, in which $\vec \varphi$ lies along the isospin z-axis at every point in 
space-time
\be
\vec \varphi=(\rho + a)\hat z,
\ee
where $\vec \rho=\rho \hat z$ is the physical excitation 
carried by $\vec \varphi$. Notice that the Lorentz-violating Chern-Simons 
term respects the SO(3) local symmetry; so after SSB occurs, we still 
have the freedom to fix the gauge, as in any normal gauge theory. 
It then turns out that the two isospin 
components of the gauge field  that are orthogonal to the chosen vacuum acquire a mass 
given by $M^2=e^2a^2$, in addition to the topological mass induced by the
Lorentz-breaking term. 
We shall make some comments on the remnant U(1)-symmetry 
along the $z$-direction latter. The new gauge action is given as below:
\bq
\Sigma_g&=&\int d^4x \left \{-\frac 14 G_{\mu \nu}^a G^{\mu \nu a} \right. \nonumber \\
&-& \left. \frac 14 \int d^4x \epsilon ^{\mu \nu \alpha \beta }c_\mu 
\left( G_{\nu \alpha}^a A_\beta^a - \frac e3 f^{abc}
A_\nu^a A_\alpha^b A_\beta^c \right) \right.\nonumber \\
&+& \left. \frac 12 M^2\left[ (A_\mu^1)^2+(A_\mu^2)^2 \right] \right\}.
\eq 
For the sake of identifying the spectrum of excitations, we can concentrate only
on the the quadratic part of the action, that is written as follows: 
\bq
\Sigma_g^{quad}&=&\int d^4x \left ( 
-\frac 14 F_{\mu \nu}^a F^{\mu \nu a} -\frac{\mu}{4}\epsilon ^{\mu \nu \alpha \beta }v_{\mu}
F_{\nu \alpha }^a A_{\beta}^a   \right. \nonumber \\
&+& \left. \frac 12 M^2\left[ (A_\mu^1)^2+(A_\mu^2)^2 \right] \right).
\eq 
In the equation above, the $F_{\mu \nu}^a$ is the usual Abelian field strength tensor
(we are not interested in the interaction Lagrangian), 
\be
F_{\mu \nu}^a = \partial_\mu A_\nu^a - \partial_\nu  A_\mu^a,
\ee
$\mu$ is a mass parameter and $v^\mu$ is an arbitrary four-vector of unit length which 
selects a preferred direction in the space-time ($c^\mu=\mu v^\mu$). This quadratic action 
can be decomposed in the sum of three actions, one for each gauge  field component:
\be
\Sigma_g^{quad}=\Sigma_g^1+\Sigma_g^2+\Sigma_g^3,
\ee
with
\bq
\Sigma_g^{1,2}&=& \int d^4x \left \{
-\frac 14 F_{\mu \nu}^{1,2} F^{\mu \nu }_{1,2} -\frac{\mu}{4}
\epsilon ^{\mu \nu \alpha \beta}v_{\mu}
F_{\nu \alpha}^{1,2}A_{\beta}^{1,2}  \right. \nonumber \\
&+& \left. \frac 12 M^2(A_\mu^{1,2})^2\right\}
\eq
and 
\be 
\Sigma_g^3= \int d^4x \left \{
-\frac 14 F_{\mu \nu}^3 F^{\mu \nu }_3 -\frac{\mu}{4}\epsilon ^{\mu \nu \alpha \beta }v_{\mu}
F_{\nu \alpha}^3A_{\beta}^3 \right \}.
\ee
The quadratic actions for the 1 and 2-components are the same as the one 
analyzed in the paper 
\cite{24} and the one referring to the 3-component has its unitarity and microcausality
investigated in the paper of reference \cite{10}. In the case of the action corresponding to 
the 3-component, the authors of \cite{3}, \cite{8}, \cite{10} study the implications on the unitarity 
and causality of the theory in the cases where, for small magnitudes, $c_\mu$ is 
time-like and space-like. The analysis shows that the behavior of these gauge 
field theories depends drastically on the space-time properties of $c_\mu$. For a purely 
space-like $c_\mu$, one finds a well-behaved Feynman propagator for the gauge field, 
and unitarity and microcausality are not violated. Contrary, a time-like $c_\mu$ 
spoils unitarity and causality.

The case of the actions for the 1- and 2-isospin components has been worked out in the paper 
\cite{24}. Like in the massless case, only for pure space-like $c_\mu$ both 
unitarity and causality can be ascertained. We can confirm these conclusions for 
the space-like $c_\mu$, by the analysis of the pole structure of the propagators. 
This is performed by the calculation of the eigenvalues of the residue matrix for each 
pole. The physical modes (particles with positive norm) have positive eigenvalues. 
The propagator is given as $\langle A_{\mu }A_{\nu }\rangle=
i({\cal O}^{-1})_{\mu \nu}$, where ${\cal O}_{\mu \nu}$ is a differential (local) 
operator identified in the action when it is written in the form
\be
\Sigma^{quad}= \frac 12\int d^4x A^\mu {\cal O}_{\mu \nu} A^\nu.
\ee 
We then have
\bq
\langle A_{\mu }A_{\nu }\rangle _3 &=&\frac{i}{D_3}\left\{ -k^{2}\theta
_{\mu \nu }-\left(- \frac {\alpha D_3}{k^2} \right. \right. \nonumber \\
&-& \left. \left. \frac{\mu^2(v\cdot k)^{2}}{
k^{2}}\right) \omega _{\mu \nu }
- i \mu S_{\mu \nu }-\mu^2\Lambda _{\mu
\nu } \right. \nonumber \\
&+& \left. \frac{\mu^2 (v\cdot k)}{k^{2}}\left( \Sigma _{\mu \nu
}+\Sigma _{\nu \mu }\right) \right\} 
\eq
and
\bq
\langle A_{\mu }A_{\nu }\rangle _{1,2}&=&\frac{i}{D_{1,2}}\left\{ -(k^{2}-M^{2})\theta
_{\mu \nu } \right. \nonumber \\
&+& \left. \left( \frac {D_{1,2}}{M^2}-\frac{\mu^2 (v\cdot k)^{2}}{%
(k^{2}-M^{2})}\right) \omega _{\mu \nu }\right.  \nonumber \\
&-&\left. i \mu S_{\mu \nu }-\frac{\mu^2k^{2}}{(k^{2}-M^{2})}\Lambda _{\mu
\nu } \right. \nonumber \\
&+& \left.\frac{\mu^2 (v\cdot k)}{(k^{2}-M^{2})}\left( \Sigma _{\mu \nu
}+\Sigma _{\nu \mu }\right) \right\},   
\label{a12}
\eq
where
\be
D_{1,2}=(k^2-m_1'^2)(k^2-m_2'^2),
\ee
\be
D_3=(k^2-m_1^2)(k^2-m_2^2),
\ee
with
\bq
m_1^2&=&\frac 12 \left [\mu^2+\mu \sqrt{\mu^2+4k_3^2}\right], \\
m_2^2&=&\frac 12 \left [\mu^2-\mu \sqrt{\mu^2+4k_3^2}\right], \\
m_1'^2&=&\frac 12 \left [2M^2+\mu^2+\mu \sqrt{\mu^2+4(M^2+k_3^2)}\right], \\
m_2'^2&=&\frac 12 \left [2M^2+\mu^2-\mu \sqrt{\mu^2+4(M^2+k_3^2)}\right].
\eq
We should say that, without loss of generality, we have fixed our external 
space-like vector as given by $v^\mu=(0,0,0,1)$. We also have chosen a 
representative momentum for $k^2>0$, given by $k^\mu=(k^0,0,0,k^3)$. 
In the propagators, we use the transverse and longitudinal spin operators, 
$\theta_{\mu \nu}$ and $\omega_{\mu \nu}$, in momenta-space,
\be
\theta _{\mu \nu }=g_{\mu \nu }-\frac{k_{\mu }k_{\nu }}{%
k^2}\;\;\mbox{and}\;\;\omega _{\mu \nu }=\frac{k_{\mu }k_{\nu }}{%
k^2},
\ee
and the $v^\mu$-dependent operators,
\be
S^{\mu \nu }=\varepsilon ^{\mu \nu \kappa \lambda }v_{\kappa }k
_{\lambda }, \,\,\, 
\Sigma _{\mu \nu }=v_{\mu }k _{\nu }\;,\;\;\lambda \equiv \Sigma
_{\mu }^{\;\mu }=v_{\mu }k^{\mu }
\ee
and
\be
\Lambda _{\mu \nu }=v_{\mu}v_{\nu }.
\ee
We then have the poles $k_0^2=k_3^2+m_1^2, \, k_3^2+m_2^2$ and 
$k_0^2=k_3^2+M^2, \,k_3^2+m_1'^2, \, k_3^2+m_2'^2$ for the propagators 
$\langle A_{\mu }A_{\nu }\rangle _3$ and $\langle A_{\mu }A_{\nu }\rangle _{1,2}$, 
respectively. 

A sensible question that arises is whether the number of degrees of freedom is maintained 
after spontaneous symmetry breaking. This is not a trivial point, since Lorentz symmetry 
is now violated. To answer it properly, we report here, as it may be found in \cite{24}, 
the calculation related to the poles of the propagators 
$\langle A_{\mu }A_{\nu }\rangle _{1,2}$ given in eq. (\ref{a12}).

With $k_{0}^{2}=k_3^2+m_1'^{2}$, we have that 
the residue matrix reads as below
\begin{widetext} 
\begin{equation}
R_{1}=\frac{1}{\sqrt{\mu ^{2}+4\left( M^{2}+k_{3}^{2}\right) }}\left( 
\begin{array}{cccc}
0 & 0 & 0 & 0 \\ 
0 & \tilde m_1^{2}-\left( M^{2}+k_{3}^{2}\right) & i\mu \tilde m_{1} & 0 \\ 
0 & -i\mu \tilde m_{1} & \tilde m_{1}^{2}-\left( M^{2}+k_{3}^{2}\right) & 0 \\ 
0 & 0 & 0 & 0
\end{array}
\right),
\end{equation}
\end{widetext}
with $\tilde m_1^2=m_1'^2+k_3^2$.
We calculate its eigenvalues and find only a single non-vanishing eigenvalue: 
\begin{equation}
\lambda =\frac{2\left| \tilde m_{1}\right| }{\sqrt{\mu ^{2}+4\left(
M^{2}+k_{3}^{2}\right) }}>0.
\end{equation}

The same procedure and the same conclusions hold through for the second zero
of $D_3$ $\left(k_{0}^{2}=k_3^2+m_{2}'^{2}\right) $: 
there comes out a unique non-vanishing eigenvalue given by $\lambda =\frac{
2\left| \tilde m_{2}\right| }{\sqrt{\mu ^{2}+4\left( M^{2}+k_{3}^{2}\right) }}
>0$.

Finally, we are left with the consideration of the pole $k_{0}^{2}=
M^{2}+k_{3}^{2}$. The residue matrix reads as follows: 
\begin{widetext}
\begin{equation}
R_{M}=\left( 
\begin{array}{cccc}
\frac{k_{3}^{2}}{M^2}  & 0 & 0 & 
\frac{\left| k_{3}\right| \left( M^{2}+k_{3}^{2}\right) ^{1/2}} {M^2}\\ 
0 & 0 & 0 & 0 \\ 
0 & 0 & 0 & 0 \\ 
\frac{\left| k_{3}\right| \left( M^{2}+k_{3}^{2}\right) ^{1/2}}{M^2} & 0 & 0 & 
\frac{\left( M^{2}+k_{3}^{2}\right)}{M^2},
\end{array}\right),
\end{equation}
\end{widetext}
and again we have obtained only a non-vanishing eigenvalue: $\lambda =\frac{1}{
M^{2}}\left( M^{2}+2k_{3}^{2}\right) >0$.

The analysis of the residue matrix at the poles yields, for each 
one, only one non-vanishing and positive eigenvalue. So, the number of degrees 
of freedom is preserved. We are before a very 
interesting result: in the ordinary mechanism, the fields $A_\mu^1$ and 
$A_\mu^2$, initially with one physical massless vector mode, and then two degrees of 
freedom, acquire, by the Higgs mechanism, a mass 
$M^2$. This pole corresponds to a mode with three degrees of freedom. 
In contrast, in the presence of our external background, there occurs 
a breakdown of this degeneracy and the Higgs mechanism 
is modified. This makes sense, since in this theory the space-time 
directions are not treated on the same footing. 
Here, the physical excitations accommodated in the fields $A_\mu^1$ and $A_\mu^2$ 
go from two massive to three massive scalar modes, as it is clear from the 
poles of eq. (\ref{a12}), given by $m_1'^2$, $m_2'^2$ and $M^2$, and each one having 
a corresponding residue matrix with a single non-trivial eigenvalue.

\section{On the existence of 't Hooft-Polyakov monopoles}

In the paper \cite{24}, it was shown that in the modified 
Electrodynamics there is no room for Dirac-like monopoles, since 
it would imply in a contradiction between two of the modified Maxwell 
equations. The effects of Chern-Simons term on 't Hooft-Polyakov monopoles have already 
been investigated in 2+1-dimensions in \cite{25b} and \cite{25c}. 
We now wish to verify whether the known solutions of 
't Hooft \cite{26} and Polyakov \cite {27} for an SO(3) theory, which 
exhibit solutions with magnetic charge, are still present whenever the external background 
is switched on.   
The model described by the action of equation (3) leads to the field 
equations 
\be
(D^2 \varphi)^a = -\lambda \varphi^a \left( |\vec \varphi|^2- a^2 \right)
\ee
and
\be
\label{31}
(D_\mu G^{\mu \alpha})^a= e f^{abc}(D^\alpha \varphi)^b \varphi^c 
+  c_\mu \tilde G^{\mu \alpha a},
\ee
where $\tilde G^{\mu \nu a}= (1/2)\epsilon^{\mu \nu \alpha \beta}G_{\alpha \beta}^a$ is the
dual tensor. We are interested in searching for static and stable solutions to these equations. At infinity, 
the energy density is given by
\bq
\label{energy}
{\cal H}&=&\frac 12 \left\{ (\vec G^{0i})^2+(-\frac 12 \epsilon _{ijk}\vec G_{ij})^2 
+(D^0 \vec \varphi)^2 + (D^i \vec \varphi)^2 \right. \nonumber \\
&+& \left. V(\vec \varphi) \right\}.
\eq
This is so because the energy-momentum tensor for the gauge field can be cast according 
to the expression below, after some algebraic relations are used:  
\bq
\Theta^{\mu \nu}&=&-G^{\mu \alpha a}G^{\nu}_{a \alpha} + \frac{g^{\mu \nu}}{4}G^{\alpha 
\beta a}G_{\alpha \beta}^a \nonumber \\
&-& \frac 14 c^\nu\epsilon^{\mu \beta \rho \sigma}
\left(G_{\beta \rho}^aA_\sigma^a-\frac {e}{3} f^{abc}A_\beta^aA_\rho^bA_\sigma^c\right),
\eq
and, when $c_\mu$ is purely space-like, as we are considering here, it 
does not contribute to the energy. 
We need a solution with finite energy and, therefore, at spatial infinity the vacuum 
conditions 
\be
G^{\mu \nu a}=  0, \,\,\,\, (D^\mu \varphi)^a=  0 
\ee
and
\be 
V(\varphi)= \frac 14 \lambda \left( |\vec \varphi|^2- a^2 \right)^2=0
\ee
must be satisfied. So, as $r \to \infty$, we must have $|\vec \varphi|=a$ and 
$\vec A_\mu =\vec 0$. The simplest non-trivial $\vec \varphi$ that satisfies 
this condition is
\be
\vec \varphi =a \hat r,
\ee
where $\hat r$ is the unit vector in the $\vec r$ direction.
This expression, that mixes a vector in the space-time and a vector
in the isospace, was called by Polyakov a "hedgehog" solution.
So, the Higgs vacuum, $(D_\mu \varphi)^a=0$, is obtained by the conventional solution,
with the most general form of $A_\mu^a$ given by
\be
\label{amu}
A_\mu^a=- \frac {1}{a^2 e} f^{abc} \varphi^b \partial  \varphi^c
+\frac 1a A_\mu  \varphi^a,
\ee
with $A_\mu$ being the projection of $\vec A_\mu$ along the direction of $\vec \varphi$
in the isospin space. The well-known field-strength tensor 
obtained by using the potential of equation (\ref{amu}) has the form
\be
\vec G_{\mu \nu}= F_{\mu \nu}\hat \varphi,
\ee
where
\be
\label{fmunu}
F_{\mu \nu}=\partial_\mu A_\nu-\partial_\nu A_\mu -
\frac {1}{ea^3} f^{abc} \varphi^a \partial_\mu \varphi^b
\partial_\nu  \varphi^c
\ee
and $\hat \varphi= \vec \varphi/a$. So, as in the conventional SO(3) theory,
in the Higgs vacuum, the unique non-vanishing component of $\vec G_{\mu \nu}$ is the one 
associated with the remnant U(1) group rotations about $\vec \varphi$. Let us see which
kind of equations the field-strength tensor satisfies in the Higgs vacuum:
\bq
\partial_\mu F^{\mu \nu}&=&\frac 1a \partial_\mu(\varphi^a G^{\mu \nu a}) \nonumber \\
&=& \frac 1a \left \{-ef^{abc}G^{\mu \nu a}A_\mu^b \varphi^c + \varphi^a\partial_\mu
G^{\mu \nu a} \right \}\nonumber \\
&=& \frac 1a \left \{ \varphi^a(D_\mu G^{\mu \nu })^a\right\}=
\frac 1a \left \{ \varphi^a c_\mu \tilde G^{\mu \nu a} \right\} \nonumber \\
&=& c_\mu \tilde F^{\mu \nu}.
\eq
In the calculation above, we have used the field equations and the condition
$(D_\mu \varphi)^a=0$. Furthemore, in a similar calculation, we can show that,
in the vacuum,
\be
\partial_\mu \tilde F^{\mu \nu}=0.
\ee
These are just the equations for the Lorentz-violating model with U(1) symmetry in
the vacuum, as studied in the papers \cite{9} and \cite{24}.

It is now important to point out some comments. In the paper \cite{24}, it was shown
that the modified Maxwell equations for the Abelian Lorentz-violating model do not
allow Dirac magnetic monopoles, for,as it is shown there, 
it would imply a contradiction between
two of the modified Maxwell equations. In the vacuum, we see that the equations now obtained are the same
and, so, this prohibition is still present. However, the new definition of the 
field-strength tensor, as given by the equation (\ref{fmunu}) (and it is due to the 
non-Abelian structure of the model), allows for a magnetic current
given by
\be
K^\mu=\partial_\nu \tilde F^{\mu \nu}=-\frac {1}{2e} \epsilon^{\mu \nu \rho \sigma}
f^{abc} \partial_\nu \hat \varphi^a \partial_\rho \hat \varphi^b \partial_\sigma \hat \varphi^c,
\ee
where $\hat \varphi^a= \varphi^a/a$. This current is identically conserved: $\partial_\mu K^\mu=0$.
The conserved magnetic charge is
\bq
\label{charge}
g&=& \int K^0 \, d^3x \nonumber \\
&=& \frac {1}{2e} \int_{S^2} \epsilon_{ijk} f^{abc}
\hat \varphi^a \partial^j \hat \varphi^b \partial^k \hat \varphi^c (d^2S)^i,
\eq
where the divergence theorem has been used and 
the integral is taken over the sphere $S^2$ at infinity. This is the boundary
of the static field configuration $\varphi$. The integral,
\be
n=\frac {1}{8\pi}
\int_{S^2} \epsilon_{ijk} f^{abc}
\hat \varphi^a \partial^j \hat \varphi^b \partial^k \hat \varphi^c (d^2S)^i,
\ee
gives the number of times $\hat \varphi(\vec r)$ covers $S^2$ as $\vec r$ runs over 
$S^2$ once. This is so because $\vec \varphi(\vec r)$ must be single-valued. So, we have
\be
\label{quant}
g=\frac{4\pi n}{e},
\ee
and we obtain the usual quantization condition. On the other hand, equation (\ref{charge}), 
by the Gauss law of the Magnetism, can be written as
\be
\label{gauss}
g=\int_{S^2} \vec B \cdot d \vec S.
\ee
In this way, we have 
\be
B^i=\frac {1}{2e}\epsilon_{ijk} f^{abc} \hat \varphi^a \partial^j 
\hat \varphi^b \partial^k \hat \varphi^c.
\ee
By using $\hat \varphi = \vec r/r$, the equation above yields 
\be
B^i=\frac 1e \frac {r^i}{r^3},
\ee 
which gives us
\be 
g=\frac {4 \pi}{e},
\ee
the minimal value compatible with the quantization condition (\ref{quant}).
We see that, at spatial infinity, in the Higgs vacuum, although 
the gauge potential $A_\mu=0$, an
electromagnetic field is contributed by the Higgs sector, and 
we get that, asymptotically, the field configurations indicate the presence 
of a non-trivial magnetic charge.  When observed 
from infinity, there is indeed a radial magnetic field.

We would like to comment on the existence of 't Hooft-Polyakov 
solutions. First of all, let us observe that, like in the Abelian case,
studied in ref. \cite{24}, the temporal 
gauge is not allowed if $\vec B \cdot \vec c \neq 0$, as can be seen 
from eq. (\ref{31}). In the Abelian case, we have been analyzing vortex-like 
solutions, and we found out that these solutions may actually show up.
Here, we are interested in a radial magnetic field 
and, so, a constant $c_\mu$ does not allow such a condition: we cannot adopt the 
temporal gauge. This is so due to the presence of an electric field along with the 
magnetic field, as it happens in the Chern-Simons theory. 
In order to verify the stability of the 't Hooft-Polyakov 
solutions, we analyze the energy density of the field configuration, 
given by eq. (\ref{energy}), written as follows:
\be
{\cal H}= {\cal H}_{temp}+\frac 12 \left((\vec G^{0i})^2+
(e\vec A^0 \times \vec \varphi)^2\right),
\ee  
where ${\cal H}_{temp}$ is the energy density we would get in the temporal gauge,
generally used in dealing with monopole calculations. This part of the energy, 
given by
\be
E_{temp}=\int d^3x\,\,{\cal H}_{temp},
\ee
can have its minimal calculated by numerical methods. There is a limiting case, 
due to Bogomol'nyi \cite{bogo}, in which a lower bound can be derived analytically, 
so that
\be
E_{temp}\geq a |g|.
\ee

In view of this, since the other terms that contribute to the energy are positive, 
we conclude that we have, also in the Lorentz-violating case, with $c_\mu$ spacelike, 
a lower bound in the energy that assures stability of the monopole configuration. 
The difference here is that the energy is contributed by the electric field too. 
However, since the temporal gauge is not possible, the solution of the field 
equations is much more complicated. In an intermediate region, even if the 
Bogomol'nyi bound is used, the usual ansatz cannot be adopted, for we do not have 
spatial isotropy anymore.
But, as stated above, we stress the appearance of a 
radial magnetic field observed in the asymptotic region. This supports the conclusion 
of the presence of a magnetic charge. However, though we do not know the full 
solution everywhere (as it happens 
for the monopoles in the situation where there is no breaking of Lorentz symmetry), 
it would be interesting to attempt at finding out the explicit solution that interpolates 
the monopole configuration from infinity to the non-asymptotic region.
That is not a trivial task and we intend to devote some time to pursue 
such an investigation.  

Another important discussion is the one related to the fact that, in the direction of 
the vacuum in isospin space, the field equations obtained are those of a U(1) 
Lorentz-breaking model with a Chern-Simons-type term. As mentioned before, these 
equations do not yield Dirac monopoles. Indeed, the presence of an explicit mass term 
(Proca-like, for instance) or the mass generation by spontaneous symmetry breaking 
always yield inconsistencies between the field equations and a Dirac-type 
monopole. This is also the case if the mass is generated by means of a Lorentz-violating 
term, as the one introduced in (\ref{2}). 
However, we see that, even if Lorentz violation occurs, there persists a solution 
in which 't Hooft-Polyakov monopoles are present. The explanation is that 
't Hooft-Polyakov monopoles are characterized by a field configuration 
set up by the triplet of Higgses. We would like to stress that the magnetic 
monopole here identified is originated from the topological character of the vacuum 
through the Higgs field, and it is actually rather different from a Dirac-like 
monopole.

\section{Concluding comments}
\indent

The purpose of this work is the investigation of two aspects 
concerning gauge theories with Lorentz- and CPT-symmetry violation: first, we analyze the 
quantization consistency of a Lorentz and CPT violating
non-Abelian model for which the group of symmetry 
is SO(3), contemporarily with the spontaneous breaking of gauge symmetry. Further, 
we investigate if the topological structure of this theory yields 't Hooft-Polyakov 
monopoles.     

The study of the quantization consistency of the theory was carried out by pursuing 
the investigation of unitarity and causality as read off from the gauge-field 
propagators. The discussion is made at the tree-approximation, without going through 
the canonical quantization procedure for field operators. We concentrate on the 
analysis of the residue matrices at each pole of the propagators. We note that we 
can split the gauge-field propagator in three parts, one for each isospin direction. 
We thus find two kinds of propagators, when spontaneous symmetry breaking takes 
place: one for the massless field $A_\mu^3$, which is in the internal direction of the 
U(1) remnant symmetry, and another for the massive, $A_\mu^1$ and $A_\mu^2$, fields.
The quantization consistency for a theory with the propagator for the massless 
$A_\mu^3$-field was analyzed in the paper \cite{3}, \cite{8}, \cite{10}. In these work, aspects like unitarity 
and microcausality were considered and the unique possibility of a consistent theory 
is the one in which the external vector, $c_\mu$, is spacelike. In the case of the fields 
$A_\mu^1$ and $A_\mu^2$, the propagator was studied in the reference \cite{24}. The 
conclusions on the $c_\mu$ vector were the same as in \cite{3}, \cite{8}, \cite{10}: it must be 
spacelike. So, the quantization of our non-Abelian theory can be carried out without 
problems provided that $c_\mu$ is spacelike.

An interesting feature of this model is the breaking of degeneracy of the degrees of 
freedom. The originally vector modes, which in Lorentz-invariant theories 
accommodate two (in the case of being massless) or three (if massive) 
degrees of freedom, are now split into two or three different 
scalar modes, as it is revealed by the eigenvalues of the residue 
matrix at the poles of the propagators. 
This is in agreement with our expectation, since we have no longer 
space-time isotropy.

We now refer to 't Hooft-Polyakov monopoles. The Lorentz-breaking term, 
with $c_\mu$ spacelike, does not interfere 
in the finiteness of the energy and we have stable solutions in the Higgs vacuum. 
The solution which yields the monopole is the usual one. An important remark must be 
pointed out that clarifies an apparent contradiction: the equations satisfied by the projection of 
the field-strength tensor in the vacuum direction are the same as the ones studied 
in \cite{24}, for which Dirac monopoles are forbidden. However, this is not the case 
here. The 't Hooft-Polyakov monopole is characterized by the configuration of 
the scalar fields. It appears as an asymptotic solution and the field configuration 
looks like a magnetic monopole when viewed from infinity.

\section* {Acknowledgments}

A. P. Baeta Scarpelli are grateful to CAPES for the invaluable financial help.

\end{document}